\documentclass[aps,prd,reprint,twocolumn,superscriptaddress,showpacs]{revtex4-1}
\usepackage{graphicx}
\usepackage{mathrsfs}
\usepackage{bm}
\usepackage{amsmath}
\usepackage{dcolumn}
\usepackage{epstopdf}
\usepackage{dsfont}
\usepackage{amssymb}
\usepackage{tabularx}
\usepackage{array}
\usepackage{float}
\usepackage{color}
\usepackage{epstopdf}
\usepackage{mathrsfs}
\usepackage[colorlinks, linkcolor=blue,anchorcolor=blue,citecolor=blue,urlcolor=blue]{hyperref}

\begin{document}
\title{Hourglass Weyl loops in two dimensions: Theory and material realization in monolayer GaTeI family}

\author{Weikang Wu}
\affiliation{Research Laboratory for Quantum Materials, Singapore University of Technology and Design, Singapore 487372, Singapore}

\author{Yalong Jiao}
\affiliation{Research Laboratory for Quantum Materials, Singapore University of Technology and Design, Singapore 487372, Singapore}

\author{Si Li}
\affiliation{Research Laboratory for Quantum Materials, Singapore University of Technology and Design, Singapore 487372, Singapore}

\author{Xian-Lei Sheng}
\affiliation{Research Laboratory for Quantum Materials, Singapore University of Technology and Design, Singapore 487372, Singapore}
\affiliation{Department of Physics, Key Laboratory of Micro-nano Measurement-Manipulation and Physics (Ministry of Education), Beihang University, Beijing 100191, China}

\author{Zhi-Ming Yu}
\affiliation{Research Laboratory for Quantum Materials, Singapore University of Technology and Design, Singapore 487372, Singapore}

\author{Shengyuan A. Yang}
\affiliation{Research Laboratory for Quantum Materials, Singapore University of Technology and Design, Singapore 487372, Singapore}

\begin{abstract}
Nodal loops in two-dimensional (2D) systems are typically vulnerable against spin-orbit coupling (SOC). Here, we explore 2D systems with a type of doubly degenerate nodal loops that are robust under SOC and feature an hourglass type dispersion.
We present symmetry conditions for realizing such hourglass Weyl loops, which involve nonsymmorphic lattice symmetries. Depending on the symmetry, the loops may exhibit different patterns in the Brillouin zone. Based on first-principles calculations, we identify the monolayer GaTeI-family materials as a realistic material platform to realize such loops. These materials host a single hourglass Weyl loop circling around a high-symmetry point. Interestingly, there is also a spin-orbit Dirac point enabled by an additional screw axis. We show that the hourglass Weyl loop and the Dirac point are robust under a variety of applied strains. By breaking the screw axis, the Dirac point can be transformed into a second Weyl loop. Furthermore, by breaking the glide mirror, the hourglass Weyl loop and the spin-orbit Dirac point can both be transformed into a pair of spin-orbit Weyl points. Our work offers guidance and realistic material candidates for exploring fascinating physics of several novel 2D emergent fermions.
\end{abstract}
\maketitle

\section{Introduction}
\label{sec_intro}
Topological materials with protected band crossings are attracting tremendous interest in current research~\cite{Bansil2016Colloquium-RoMP,Chiu2016Classification-RoMP,Burkov2016Topological-Nm,Yan2017Topological-ARoCMP,Armitage2018Weyl-RoMP}. The study was initiated by drawing insightful analogy between fundamental particles in the relativistic quantum field theory and low-energy emergent fermions in condensed matters. In this way, the Weyl and Dirac semimetals were discovered~\cite{Murakami2007Phase-NJoP,Wan2011Topological-PRB,Young2012Dirac-Prl,Wang2012Dirac-PRB,Wang2013Three-PRB,Yang2016Dirac-spin,Armitage2018Weyl-RoMP}, which have twofold and fourfold degenerate band crossing points, and around these points, the low-energy electrons resemble the Weyl and Dirac fermions and can exhibit fascinating physical effects like their counterparts in high energy physics~\cite{Nielsen1983Adler-PLB,Volovik2003universe-OUPoD,Guan2017Artificial-nQM}. Moving forward, it was realized that crystalline solids may host more types of emergent fermions beyond the Weyl/Dirac paradigm~\cite{Bradlyn2016Dirac-S,Bzdusek2016Nodal-N,Wang2016Hourglass-N,Wang2017Hourglass-Nc,Hu2019Three-C}. For example, in a three-dimensional (3D) material, besides 0D nodal point, band crossings may also take the form of 1D nodal loops~\cite{Yang2014Dirac-Prl,Weng2015Topological-PRB,Mullen2015Line-Prl,Yu2015Topological-Prl,Kim2015Dirac-Prl,Chen2015Nanostructured-Nl,Fang2016Topological-CPB,Li2017Type-PRB,Yu2019Quadratic-PRB,Chen2019Weyl-PRB} or even 2D nodal surfaces~\cite{Liang2016Node-PRB,Zhong2016Towards-N,Wu2018Nodal-PRB,Zhang2018Nodal-PRB,Gao2019Hexagonal-PRM}.

Quantum energy levels tend to repel each other, so the band crossings in topological materials generally require topology or symmetry protections.
The following two factors play important roles regarding the stability of band crossings. The first is the dimensionality of the material system. Three-dimensional materials can have more varieties of topological band crossings with more symmetry protections. In comparison, the condition for protected band crossings in 2D materials is much more stringent, due to the reduced number of symmetries. This is clearly reflected in the number of space groups: 230 for 3D versus 80 (layer groups) for 2D~\cite{Litvin1991Character-Plenum}. The second is the effect of spin-orbit coupling (SOC). In the absence of SOC, electrons can be regarded as spinless and the crystalline symmetries are described by single valued representations. However, when SOC is taken into account, the symmetries must be described by double valued representations. Generally, the number of double valued representations for a space group is less than the number of single valued ones. It follows that a band crossing that is protected in the absence of SOC often gets destroyed when SOC is turned on.

From the above considerations, one can see that it is a challenging task to find 2D topological materials, and it is even more challenging to require the band crossings in such 2D materials to be robust against SOC. As an example, the ``Dirac points" in most 2D materials including graphene are in fact not stable under SOC. Young and Kane~\cite{Young2015Dirac-Prl} pointed out that a truly stable Dirac point in 2D requires certain nonsymmorphic symmetry protection, and the first realistic 2D material system with such spin-orbit Dirac points was found by Guan \emph{et al.}~\cite{Guan2017Two-PRM} in monolayer HfGeTe-family materials. Here, it should be mentioned that an additional difficulty in finding 2D topological materials comes from the structural stability: Existing 2D materials respecting the particular symmetry requirement are quite limited, while materials artificially constructed to follow the symmetry requirement are often structurally unstable.

For 2D systems, the topological band crossings may also take the form of 1D nodal loops~\cite{Feng2017Experimental-Nc,Gao2018Epitaxial-AM,Zhong2019Two-N,Wang2019Two-apa}. Motivated by the recent works on spin-orbit Dirac points in 2D~\cite{Young2015Dirac-Prl,Guan2017Two-PRM}, one may naturally wonder: \emph{Can we also find stable 2D materials with nodal loops robust against SOC?} In Ref.~\cite{Young2015Dirac-Prl}, a kind of spin-orbit nodal loop in 2D was noticed from a model study. However, the symmetry condition has not been clarified, and a realistic 2D material hosting such kind of loops has not been found yet. In this paper, we reveal the first realistic 2D material with a nodal loop robust under SOC. The nodal loop we find here is twofold degenerate and features an hourglass-type dispersion, hence is termed as a hourglass Weyl loop.
We clarify the symmetry conditions for realizing hourglass Weyl loops in 2D systems. We show that depending the symmetry, the hourglass Weyl loops may exhibit different patterns in the Brillouin zone (BZ).
Based on first-principles calculations, we identify the monolayer GaTeI family materials, which can be readily exfoliated from their existing 3D bulk crystals, as a realistic 2D platform hosting hourglass Weyl loops. Besides, the system also hosts a spin-orbit Dirac point. We show that breaking symmetry can transform the band crossings into other interesting types: Breaking a screw rotation can generate a pair of hourglass Weyl loops, and breaking a glide mirror can generate two pairs of spin-orbit Weyl points. We emphasize that the study here is on 2D systems, distinct from the works on 3D. The hourglass Weyl loop represents a new type of nodal loops, and it is different from those concepts defined in terms of dispersions (such as type-II~\cite{Li2017Type-PRB} or hybrid loops~\cite{Li2017Type-PRB,Zhang2018Hybrid-PRB}).
Note that the spin-orbit Weyl point here is doubly degenerate and is robust under SOC, hence it is \emph{distinct} from the Dirac points or the spin-orbit-free Weyl points discussed before. Our work offers guidance and a concrete material platform for exploring the fundamental physics of spin-orbit Weyl loops and Weyl points in 2D systems.

\section{Symmetry analysis for hourglass Weyl loops in 2D}\label{sec_symm_hg_loop}
\begin{figure*}
	\centering
	\includegraphics[width=0.8\textwidth]{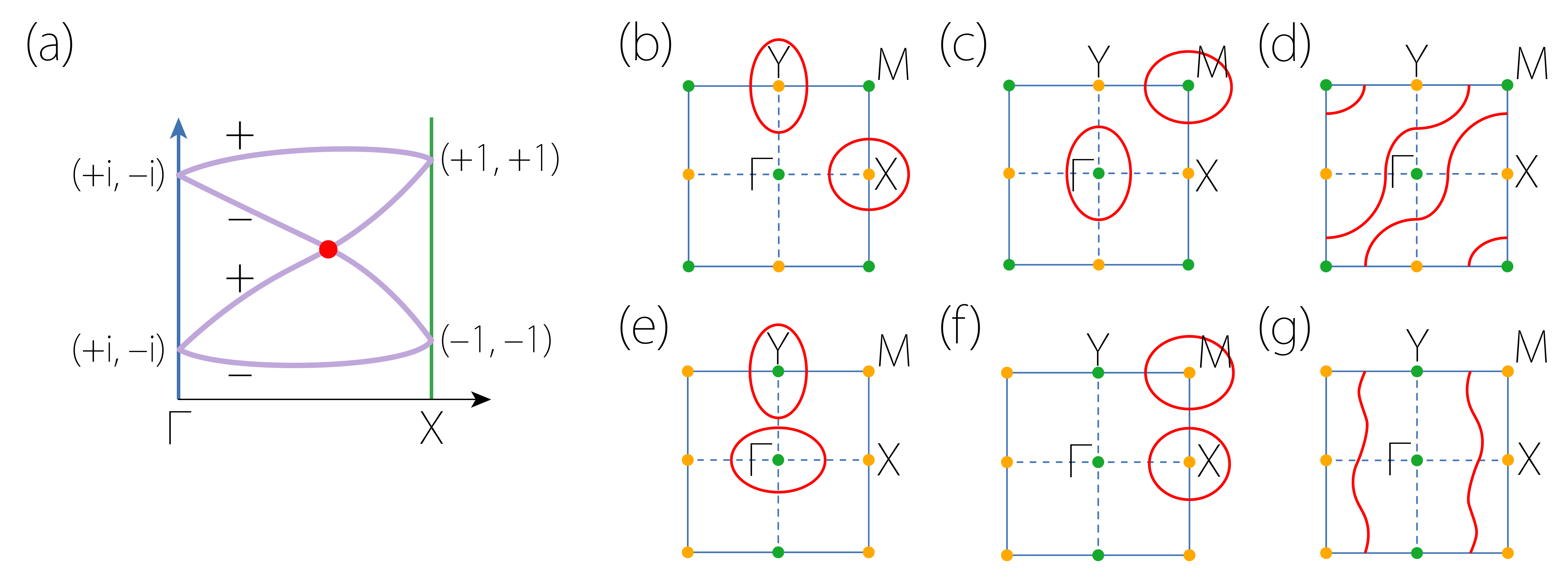}
	\caption{(a) Schematic figure showing the hourglass dispersion along a path connecting $\Gamma$ and $X$. The labels indicate the $\widetilde{M}_{z}$ eigenvalues. Partner switching between two doublets leads to the twofold Weyl crossing point (red dot). (b-d) Possible patterns of Weyl loops in the 2D BZ for the glide mirror $\{M_{z}|\frac{1}{2}\frac{1}{2}\}$. (e-g) Possible patterns of Weyl loops for the glide mirror $\{M_{z}|\frac{1}{2}0\}$. In panels (e-g), the yellow and green dots denote the TRIM-A and the TRIM-B points, respectively.}
	\label{fig_hourglass}
\end{figure*}

Let us start by analyzing the hourglass Weyl loops from a symmetry perspective. We consider a 2D nonmagnetic system with non-negligible SOC. Hence, the time reversal symmetry $\mathcal{T}$ is preserved, and $\mathcal{T}^2=-1$ as for a spinful system. By definition, a Weyl loop is doubly degenerate. This indicates that the inversion symmetry $\mathcal{P}$ must be broken, otherwise each band would be doubly degenerate due to the $\mathcal{PT}$ symmetry and their crossings would be at least fourfold degenerate.

To enable the presence of hourglass Weyl loops, we find that one minimal set of symmetries is $\mathcal{T}$ plus a nonsymmorphic glide mirror $\widetilde{\mathcal{M}}_{z}=\{M_{z}|\bm{t}_{\parallel}\}$ where $\bm{t}_{\parallel}$ is a half lattice translation parallel to the mirror plane (we assume the 2D material lies in the $x$-$y$ plane). The nodal loop pattern may depend on the direction of this half lattice translation. Below, we shall discuss two cases of the glide mirror one by one: (i)
$\{M_{z}|\frac{1}{2}\frac{1}{2}\}$; and (ii) $\{M_{z}|\frac{1}{2}0\}$ or $\{M_{z}|0\frac{1}{2}\}$.

\emph{Case-I: $\widetilde{\mathcal{M}}_{z}=\{M_{z}|\frac{1}{2}\frac{1}{2}\}$.}

Each $k$ point in the 2D BZ is invariant under $\widetilde{\mathcal{M}}_{z}$, so any Bloch state $|u\rangle$ at $k$ can be chosen as eigenstate of $\widetilde{\mathcal{M}}_{z}$. In the presence of SOC, one finds that
\begin{equation}\label{glide1}
  (\widetilde{\mathcal{M}}_{z})^{2}=T_{11}\bar{E}=-e^{-ik_{x}-ik_{y}},
\end{equation}
where $T_{11}$ denotes the translation along both $x$ and $y$ directions by a lattice constant, and $\bar{E}$ is the $2\pi$ spin rotation. Hence, the eigenvalues of $\widetilde{\mathcal{M}}_{z}$ are given by
\begin{equation}
  g_{z} =\pm ie^{-ik_{x}/2-ik_{y}/2},
\end{equation}
which are $k$-dependent.

There are four time reversal invariant momenta (TRIM) points in the 2D BZ, labeled as $\Gamma$, $X$, $Y$ and $M$, as illustrated in Fig.~\ref{fig_hourglass}(b). At these points, the bands must form degenerate Kramers pairs due to the presence of $\mathcal{T}$. Let's consider the $\widetilde{\mathcal{M}}_{z}$ eigenvalues $g_z$ at these points. For example, at $\Gamma$, we have $g_z=\pm i$, so each Kramers pair $|u\rangle$ and $\mathcal{T}|u\rangle$ must have opposite $g_z$. However, at the $X$ point $(\pi, 0)$, since $g_{z}=\pm 1$, each Kramers pair $|u\rangle$ and $\mathcal{T}|u\rangle$ must share the same $g_{z}$. Due to this different pairing at $\Gamma$ and $X$, there must be a switch of partners between two pairs when going from $\Gamma$ to $X$, during which the four bands must be entangled to form the hourglass dispersion, as schematically shown in Fig.~\ref{fig_hourglass}(a). Previous works have shown that similar hourglass-type dispersion may also appear on the surface of a 3D system, e.g., on the (010) surface of the 3D insulator KHgSb~\cite{Wang2016Hourglass-N,Ma2017Experimental-Sa,Liu2017Finite-PRB}. Notably, the doubly degenerate neck crossing point of the hourglass is protected, because the two crossing bands have opposite $g_{z}$.

The above discussion suggests that we can classify the TRIM points into two types based on the value of $g_z$: type-A with $g_z=\pm 1$ and type-B with $g_z=\pm i$. For the current case, $X$ and $Y$ belong to TRIM-A, whereas $\Gamma$ and $M$ belong to TRIM-B, as illustrated in Figs.~\ref{fig_hourglass}(b-d) by using different colors. The key point is that when going from any TRIM-A point to any TRIM-B point along an \emph{arbitrary} path, there must be an hourglass spectrum and hence must exist a Weyl crossing point on this path. The path can be arbitrary, because any $k$ point in the BZ is invariant under the glide mirror. Therefore, these crossing points must trace out hourglass Weyl loops in the 2D BZ.

Further taking into account the constraint of the $\mathcal{T}$ symmetry, we find that the hourglass Weyl loops can assume the patterns as illustrated in Figs.~\ref{fig_hourglass}(b)-\ref{fig_hourglass}(d). In Fig.~\ref{fig_hourglass}(b) and \ref{fig_hourglass}(c), we have localized loops circling around two out of the four TRIM points. In Fig.~\ref{fig_hourglass}(d), we have two extended loops, each traversing the BZ. It was pointed out by Li \emph{et al.}~\cite{Li2017Type-PRB} that a localized nodal loop circling around a high-symmetry point is topologically distinct from an extended loop traversing the BZ, according to their winding patterns around the BZ. For example, each loop in Fig.~\ref{fig_hourglass}(d) winds around the BZ by one time in each direction, so it may be assigned a topological index of $(1,1)$. In contrast, each localized loop in Fig.~\ref{fig_hourglass}(b)
and Fig.~\ref{fig_hourglass}(c) does not wind around the BZ, so it is characterized by an index of $(0,0)$.

\emph{Case-II: $\widetilde{\mathcal{M}}_{z}=\{M_{z}|\frac{1}{2}0\}$ or $\{M_{z}|0\frac{1}{2}\}$.}

For concreteness, let us consider $\widetilde{\mathcal{M}}_{z} = \{M_{z}|\frac{1}{2}0\}$ in the following discussion, and the case for $\widetilde{\mathcal{M}}_{z} = \{M_{z}|0\frac{1}{2}\}$ will be similar. We have
\begin{equation}\label{glide2}
(\widetilde{\mathcal{M}}_{z})^{2}=T_{10}\bar{E}=-e^{-ik_{x}}.
\end{equation}
Hence the eigenvalues of $\widetilde{\mathcal{M}}_{z}$ are given by $g_{z} =\pm ie^{-ik_{x}/2}$, depending only on $k_{x}$. Compared with Case-I, the main difference here is that the types of the TRIM points are changed. Now, $X$ and $M$ are TRIM-A points, whereas $\Gamma$ and $Y$ are TRIM-B points, as illustrated in Figs.~\ref{fig_hourglass}(e)-\ref{fig_hourglass}(g). Repeating the analysis as we did for Case-I, one finds that hourglass Weyl loops also exist for the current case, and the possible patterns are shown in Figs.~\ref{fig_hourglass}(e)-\ref{fig_hourglass}(g).
This finishes our symmetry analysis.

We have several comments before proceeding. First, the hourglass Weyl loops resulting from the above symmetry analysis are \emph{essential} band crossings, which means that they are guaranteed to appear in the band structure as long as the symmetry condition is satisfied.

Second, we have mentioned that in order to have a doubly degenerate Weyl loop, the inversion symmetry $\mathcal{P}$ must be broken. Since $\mathcal{PT}$ leads a double degeneracy for each state, one may ask whether including $\mathcal{P}$ would naturally double the Weyl loop, i.e., making the Weyl loop into a fourfold degenerate Dirac loop? The answer is negative. This is because for an eigenstate $|u\rangle$ of $\widetilde{\mathcal{M}}_{z}$ (either in Case-I or Case-II) with an eigenvalue $g_{z}$, one can show that
\begin{equation}
  \widetilde{\mathcal{M}}_{z}(\mathcal{P}\mathcal{T}|u\rangle)=-g_{z}(\mathcal{P}\mathcal{T}|u\rangle).
\end{equation}
Hence, the degenerate pair $|u\rangle$ and $\mathcal{P}\mathcal{T}|u\rangle$ at each $k$ point have opposite $g_{z}$. This rules out the neck crossing point in the hourglass, because the crossing bands do not have distinct $\widetilde{\mathcal{M}}_{z}$ eigenvalues. Thus, a Dirac loop cannot be stabilized in such case. The symmetry conditions for stabilizing hourglass Dirac loops in \emph{3D systems} can be found in Ref.~\cite{Wang2017Hourglass-Nc,Li2018Nonsymmorphic-PRB}.

Third, the topological band features studied here are for the \emph{bulk} states of a 2D system. For a 2D system with length scales much larger than the lattice spacing, the result must be insensitive to the boundary conditions. However, if the system is quantum confined along certain direction, then the band structure would be quantized into subbands and become much different. A work by Araujo \emph{et al.}~\cite{Araujo2016Topological-PRB} showed that the dispersion for the topological edge states in a nanoribbon could sensitively depend on the edge configuration. Here, when a 2D hourglass-Weyl-loop system is made into nanoribbons, the presence of hourglass-type dispersion in the ribbon band structure would depend on whether the relevant glide mirror is still preserved in the ribbon geometry or not.

Finally, we assumed no other symmetry in the above analysis. Additional symmetries may lead to additional degeneracies, and may affect the hourglass Weyl loops. As we will see in the following section, the presence of an additional screw axis may transform one of the Weyl loops into a spin-orbit Dirac point.

\section{Monolayer GaTeI family}
\label{sec_hg_GaTeI}

As derived in the previous section, the requirements to have an hourglass Weyl loop are: (i) nonnegligible SOC; (ii) preserved $\mathcal{T}$ and $\widetilde{\mathcal{M}}_z$ symmetries; and (iii) absence of $\mathcal{P}$.
Guided by these, we find a concrete material realization---the monolayer GaTeI family materials.

\subsection{Structure and Stability}

\begin{figure}
	\centering
	\includegraphics[width=0.49\textwidth]{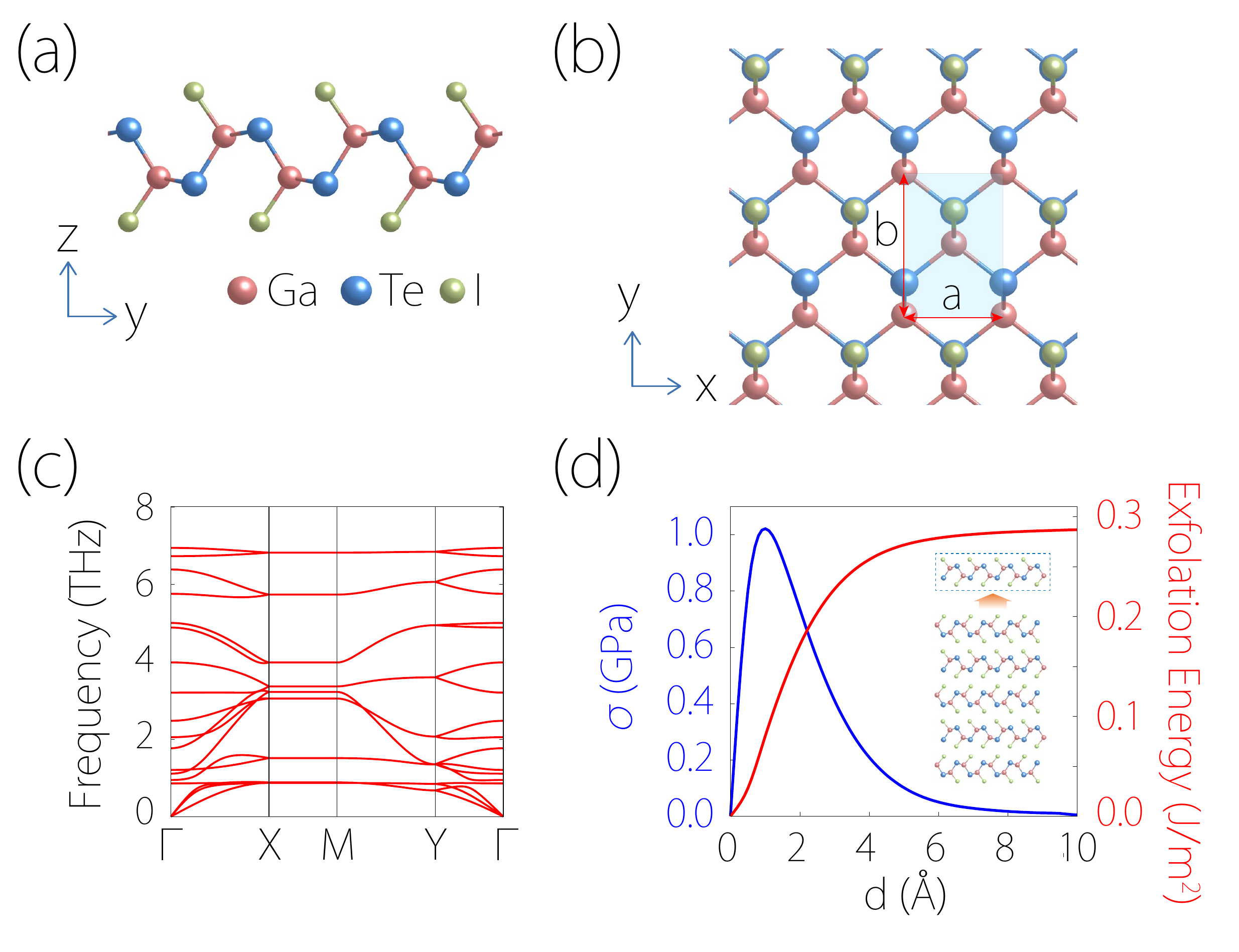}
	\caption{(a) Side and (b) top view of the crystal structure of monolayer GaTeI. The blue shaded region in (b) indicates the unit cell. $a$ and $b$ are the lattice parameters. (c) Phonon spectrum for the monolayer, showing the dynamical stability of the structure. (d) Calculated exfoliation energy (red line) for monolayer GaTeI as a function of the separation distance $d$ from the bulk (as shown in the inset). Here the bulk is modeled by six GaTeI layers in the calculation. The blue curve shows the exfoliation strength $\sigma$ (i.e., the derivative of exfoliation energy with respect to $d$).}
	\label{fig_struc}
\end{figure}

The 3D bulk of the Ga$XY$ ($X =$ Se, Te; $Y =$ Cl, Br, I) compounds, as an intermediate phase of the system Ga$Y_{3}$-Ga$_{2}X_{3}$, were synthesized experimentally in the 1980s~\cite{Wilms1981GaTeCl-ZfuNB,Kniep1983Phase-MRB}. These bulk compounds share a layered black-phosphorus-type structure with AB stacking~\cite{Kniep1983Phase-MRB}. The structure for a monolayer of these compounds is shown in Figs.~\ref{fig_struc}(a) and \ref{fig_struc}(b). One notes that although the bulk materials have inversion symmetry with the space group $Pnnm$ (No.~58), their monolayers possess $Pmn2_{1}$ (No.~31) symmetry where the inversion symmetry is violated. Some member of this family such as the GaTeCl monolayer was predicted to be an indirect band-gap semiconductor with robust ferroelasticity and ferroelectricity in recent studies~\cite{Zhang2018controllable-N,Zhou2018DFT-N}. In this work, since we want a strong SOC strength, the monolayer with heavier elements such as Te and I is desired. In the following discussion, we focus on the monolayer GaTeI as a representative example. The results for other members are mentioned in the discussion section.

The structure for monolayer GaTeI can be considered as a GaTe monolayer (isostructural to black phosphorene) surface functionalized by iodine atoms which connect to Ga atoms. Thus, it forms a sandwich structure in the sequence of I-GaTe-I, and the unit cell is of a rectangular lattice. The lattice parameters for the 3D bulk form of GaTeI obtained from our DFT calculations (calculation methods are presented in the Appendix~\ref{asec_DFT}) are given by $a = 4.147\ \text{\AA}$, $b = 6.170\ \text{\AA}$, and $c = 16.092 \ \text{\AA}$; while for the monolayer, $a = 4.235\ \text{\AA}$ and $b = 6.098\ \text{\AA}$.

To check the structural stability of monolayer GaTeI, we calculate its phonon spectrum. As observed from Fig.~\ref{fig_struc}(c), there is no imaginary frequency (soft mode) throughout the BZ, which indicates the dynamical stability of the material.

Next, we check the possibility to exfoliate one monolayer GaTeI from its 3D bulk sample. We calculate the exfoliation energy and the exfoliation strength [see Fig.~\ref{fig_struc}(d)]. The exfoliation strength $\sigma$ is obtained as the maximum derivative of exfoliation energy with respect to the separation distance $d$. For bulk samples, the binding between GaTeI layers is relatively weak. With increasing $d$, the energy quickly saturates to a value corresponding to the exfoliation energy at about 0.29~J/m$^{2}$, and the maximum exfoliation strength is about 1.0~GPa. They are even less than those of graphene (0.37~J/m$^{2}$ and 2.1~GPa)\cite{Zacharia2004Interlayer-PRB} and Ca$_{2}$N (1.14~J/m$^{2}$ and 4.42~GPa)~\cite{Zhao2014Obtaining-JotACS,Guan2015Electronic-Sr}, suggesting the feasibility to obtain monolayer GaTeI by mechanical exfoliation method.

\subsection{Hourglass Weyl Loop and Spin-Orbit Dirac Point}
\label{subsec_hg_Dirac}
The GaTeI monolayer possesses the nonsymmorphic space group $Pmn2_{1}$ (No.~31), which can be generated by the following two symmetry elements: the glide mirror $\widetilde{\mathcal{M}}_{z}$:~$(x,y,z)\rightarrow(x+\frac{1}{2},y+\frac{1}{2},-z)$ and the screw axis $\widetilde{\mathcal{C}}_{2y}$:~$(x,y,z)\rightarrow(-x+\frac{1}{2},y+\frac{1}{2},-z)$. Besides, no magnetic ordering has been found for the material, so the time reversal symmetry $\mathcal{T}$ is also preserved. Therefore, monolayer GaTeI satisfies the conditions for hourglass Weyl loops listed at the beginning of Sec.~\ref{sec_hg_GaTeI}.

\begin{figure}
\centering
\includegraphics[width=0.49\textwidth]{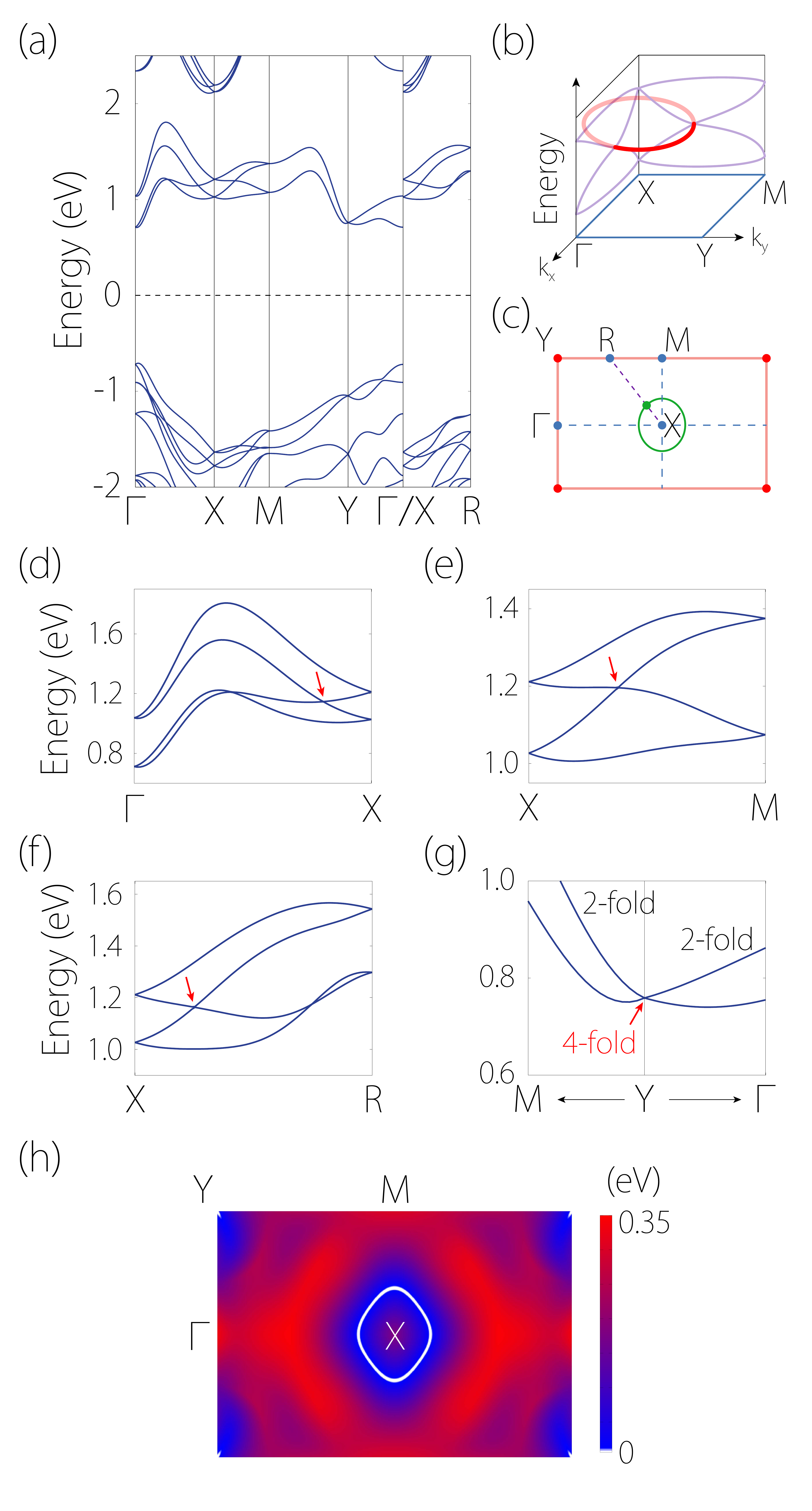}
\caption{(a) Electronic band structure of the monolayer GaTeI with SOC included. The Fermi energy is set at the middle of the gap. (b) Schematic figure showing the hourglass Weyl loop around $X$. An arbitrary path connecting $X$ and any point on the $k_{x}=0$ or $k_{y}=\pm \pi$ line will exhibit the hourglass spectrum. (c) 2D Brillouin zone with high symmetry points labeled. The green loop around $X$ indicates the hourglass Weyl loop. The red-colored boundaries are the lines with two-fold band degeneracy. An essential Dirac point is located at point $Y$, as indicated by the red dot. Panels (d-f) show the zoom-in images for the band dispersion along $\Gamma$-$X$, $X$-$M$, and $X$-$R$. Here $R$ is the midpoint between $M$ and $Y$. The red arrows indicate the twofold Weyl crossing points. (g) Enlarged band structure around $Y$. The red arrow indicates the spin-orbit Dirac point. (h) Shape of the hourglass Weyl loop (white-colored loop) obtained from the DFT calculations. The color map corresponds to the energy difference between the two crossing bands.}
\label{fig_bands}
\end{figure}

Figure~\ref{fig_bands}(a) shows the calculated band structure of monolayer GaTeI with SOC included. One finds that the material is a semiconductor with a band gap of 1.42~eV, with its valence band maximum (VBM) and conduction band minimum (CBM) both located near $\Gamma$. 
In comparison with the band structure without SOC (see Fig.~\ref{fig_bands_noSOC}), one can see that the SOC strongly modifies the degeneracy and band crossings in the band structure. Take the bands near CBM as an example. The nodal line along $X$-$M$ is removed, but there emerges an interesting hourglass-type dispersion on $X$-$M$ [see Fig.~\ref{fig_bands}(e) for a zoom-in image]. Such a hourglass dispersion also appears on the path $\Gamma$-$X$.  According to the symmetry analysis in Sec.~\ref{sec_symm_hg_loop}, due to $\widetilde{\mathcal{M}}_{z}$ and $\mathcal{T}$, the neck crossing points of the hourglass dispersion on these paths should trace out an hourglass Weyl loop. We pick a generic path connecting $X$ to a point $R$ on $Y$-$M$, and also find an hourglass dispersion on $X$-$R$ as shown in Fig.~\ref{fig_bands}(f). We scan the BZ and plot the shape of this loop as shown in Fig.~\ref{fig_bands}(h). The calculation indeed confirms the existence of a Weyl loop around $X$ in the band structure of monolayer GaTeI.


Since $\widetilde{\mathcal{M}}_{z}$ involves half lattice translations along both $x$ and $y$ directions, one may expect a second loop around the $Y$ point [corresponding to the case in Fig.~\ref{fig_hourglass}(b)]. However, we do not see such a loop in the band structure shown in Figure~\ref{fig_bands}(a). Instead, there is a fourfold band degeneracy at $Y$ [see Fig.~\ref{fig_bands}(g)], corresponding to a spin-orbit Dirac point. Why is this?
The reason is due to the presence of the extra screw rotation $\widetilde{\mathcal{C}}_{2y}$, which shrinks the hourglass Weyl loop into a Dirac point at $Y$.
Specifically, at $Y$ ($0,\pi$), we have the $\widetilde{\mathcal{M}}_{z}$ eigenvalue $g_{z}=\pm1$, and the states can be chosen as the $\widetilde{\mathcal{M}}_{z}$ eigenstates, labeled by $|g_z\rangle$. As shown in the Appendix~\ref{asec_WeylGY},  $\widetilde{\mathcal{M}}_{z}$ and $\widetilde{\mathcal{C}}_{2y}$ anticommute at $Y$, so the energy eigenstates $|g_{z}\rangle$ and $\widetilde{\mathcal{C}}_{2y}|g_{z}\rangle$ at $Y$ must have opposite $\widetilde{\mathcal{M}}_{z}$ eigenvalues, one with $g_{z}=+1$ and the other with $g_{z}=-1$. Besides, $Y$ is also a TRIM point, and thus any states $|g_{z}\rangle$ at $Y$ have another degenerate Kramers partner $\mathcal{T}|g_{z}\rangle$ with the same $\widetilde{\mathcal{M}}_{z}$ eigenvalues ($+1$ or $-1$). This ensures fourfold degeneracy for any state at $Y$, with the following four linearly independent states $\{|g_{z}\rangle,\widetilde{\mathcal{C}}_{2y}|g_{z}\rangle,\mathcal{T}|g_{z}\rangle,\mathcal{T}\widetilde{\mathcal{C}}_{2y}|g_{z}\rangle\}$. Note that for a generic $k$ point deviating from $Y$, it will not be invariant under $\mathcal{T}$, hence the fourfold degeneracy will generally be lifted away from $Y$. Thus, an isolated Dirac point must appear at $Y$. Such Dirac point is robust against SOC, similar to the one found in monolayer HfGeTe~\cite{Guan2017Two-PRM}.

In addition, we also note that the bands along the paths $\Gamma$-$Y$ and $Y$-$M$ are doubly degenerate, forming Weyl nodal lines [see Fig.~\ref{fig_bands}(g)]. The degeneracy on $\Gamma$-$Y$ is due to the anticommutation relation $\{\widetilde{\mathcal{C}}_{2y}, \widetilde{\mathcal{M}}_{z}\} = 0$ on this path, while the degeneracy on $Y$-$M$ is due to the $\mathcal{T}\widetilde{\mathcal{C}}_{2y}$ symmetry which satisfies $(\mathcal{T}\widetilde{\mathcal{C}}_{2y})^{2}=-1$ on this path. The detailed analysis on these is presented in the Appendix~\ref{asec_WeylGY} and~\ref{asec_WeylYM}. Furthermore, since the doubly degenerate Bloch states along $\Gamma$-$Y$ and $Y$-$M$ evolve from $\Gamma$ and $M$, respectively, they should have opposite $\widetilde{\mathcal{M}}_{z}$ eigenvalues, similar to the TRIM-B points. Hence, the symmetry argument in Sec.~\ref{sec_symm_hg_loop} can be extended to apply for an arbitrary path connecting $X$ (a TRIM-A point) and a point on the $\Gamma$-$Y$ or $Y$-$M$ ($Y$ should be excluded due to the fourfold degeneracy for any state at $Y$), which must feature an hourglass dispersion with a doubly degenerate neck Weyl crossing point. These crossing points form the hourglass Weyl loop centered around $X$.

\subsection{Strain Effects on Band Crossings}

\begin{figure}
	\includegraphics[width=0.5\textwidth]{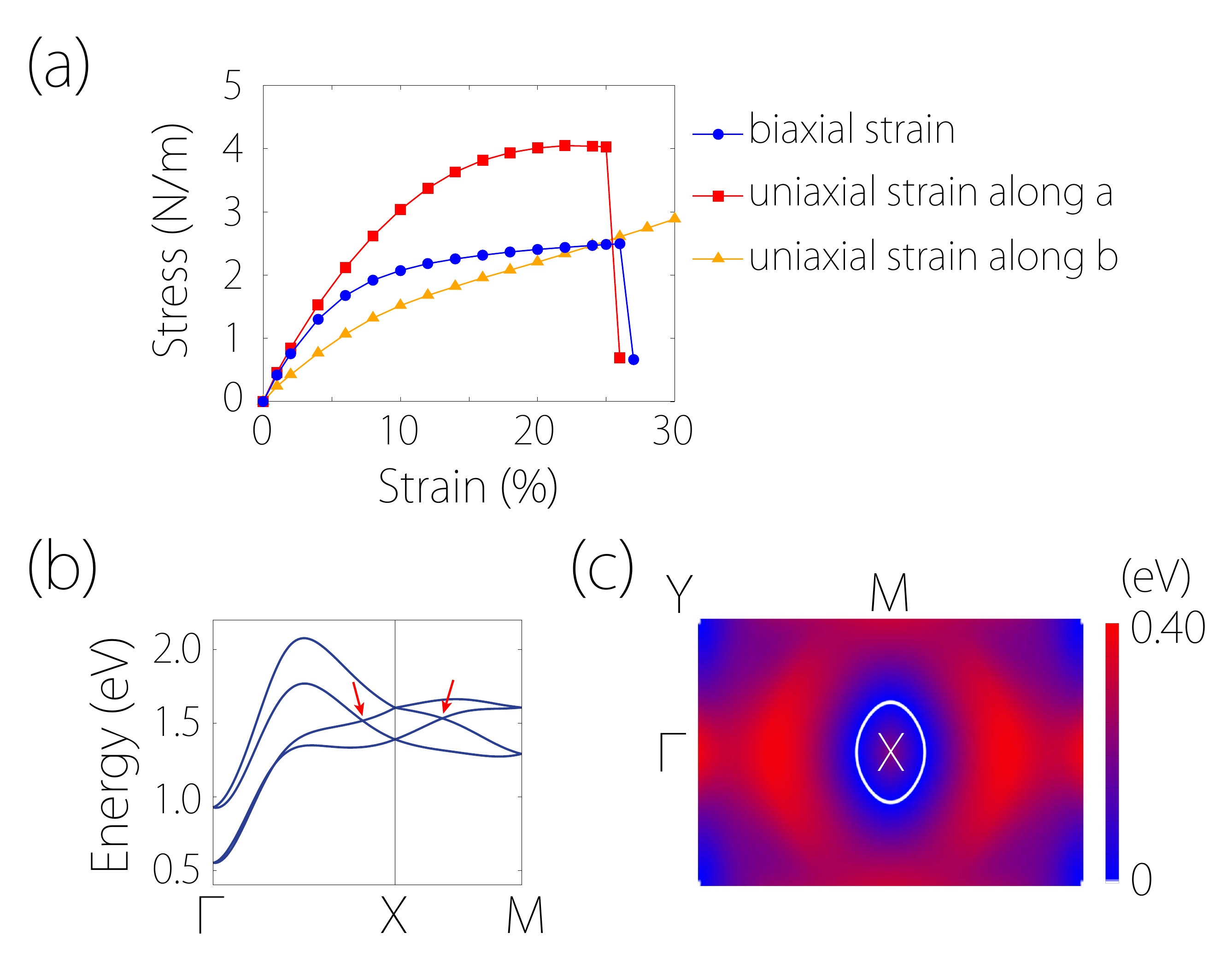}
	\caption{(a) Stress-strain relations for monolayer GaTeI under different types of strain. (b) Zoom-in image around the hourglass Weyl loop under $+6\%$ biaxial strain. The red arrows indicate the neck crossing points which trace out the loop around $X$. Panel (c) shows the shape of the hourglass Weyl loop (white-colored loop) under $+6\%$ biaxial strain. The color map indicates the energy difference between the two crossing bands.}
	\label{fig_strain}
\end{figure}

2D materials typically have good mechanical properties. In Fig.~\ref{fig_strain}(a), we plot the calculated strain-stress curves for monolayer GaTeI. It shows a linear elastic region up to $8\%$ strain, and the critical strain is more than $20\%$, suggesting that strain can be employed as an effective way to tune the properties of monolayer GaTeI.

Since the hourglass Weyl loop around $X$ is protected by the $\mathcal{T}$ and $\widetilde{\mathcal{M}}_{z}$ symmetries, it cannot be destroyed as long as these two symmetries are preserved. We find that these symmetries can survive under a variety of strains, such as in-plane biaxial, uniaxial and shear strains. In Fig.~\ref{fig_strain}(b), we plot the calculated band structures of monolayer GaTeI under the $+6\%$ biaxial strain as an example. One indeed observes that the hourglass Weyl loop retains, only the shape and the energy of the loop changed by strain.

\begin{figure}
	\includegraphics[width=0.45\textwidth]{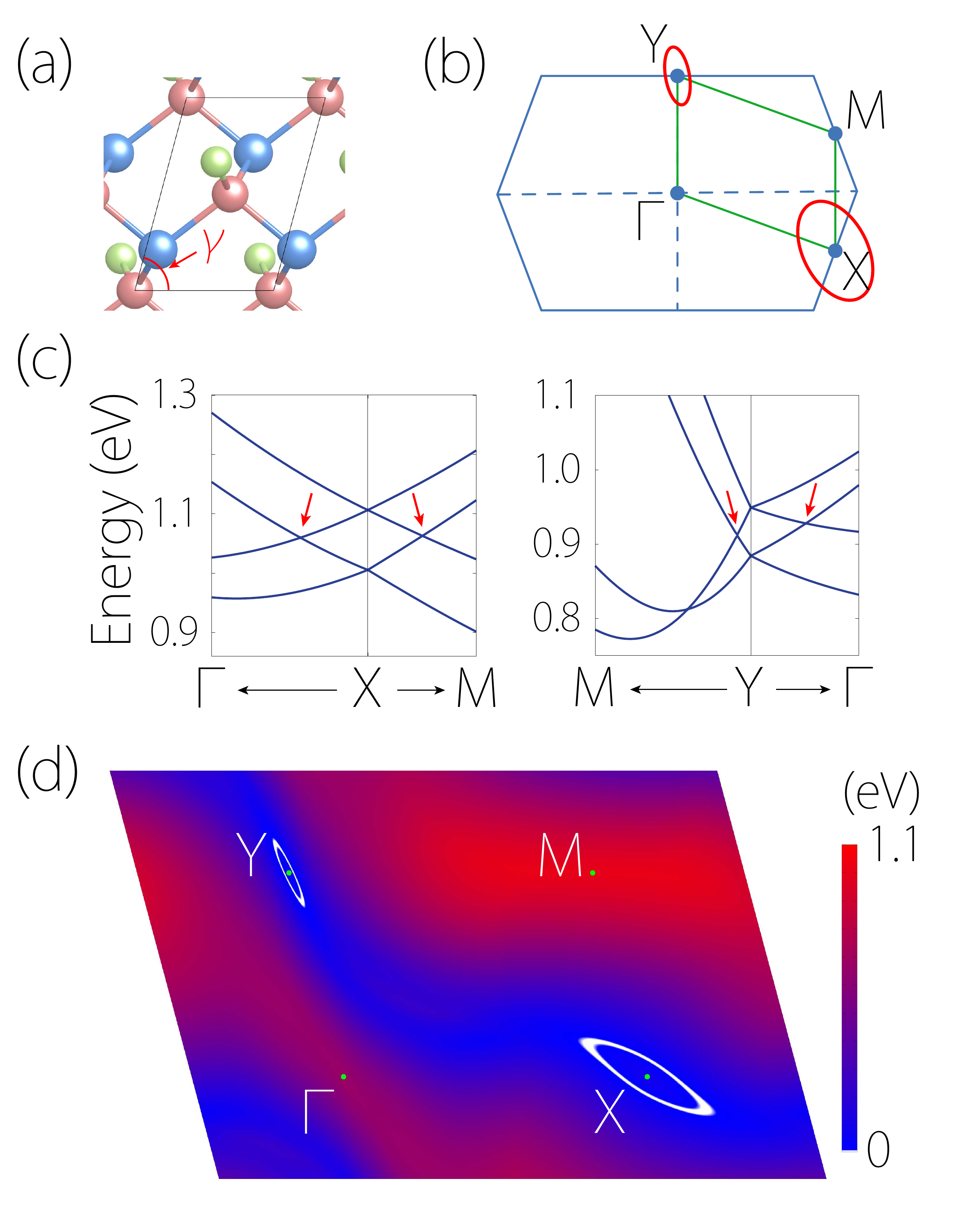}
	\caption{(a) Lattice distortion of monolayer GaTeI by varying the angle $\gamma$ between $a$ and $b$. (b) 2D Brillouin zone for the distorted lattice. The red loop around $X$ and $Y$ indicate the two hourglass Weyl loops schematically. Panel (c) shows the zoom-in images for the bands along $\Gamma$-$X$-$M$ and $M$-$Y$-$\Gamma$, with $\gamma = 75^{\circ}$. The red arrows indicate the neck crossing points on the two hourglass Weyl loops. (d) Shape of the hourglass Weyl loop (white-colored) when $\gamma = 75^{\circ}$. The color map indicates the energy difference between the two crossing bands. The green dots indicate the high-symmetry points.}
	\label{fig_shear}
\end{figure}

The shear strain can change the type of Bravais lattice for monolayer GaTeI. It preserves $\widetilde{\mathcal{M}}_{z}$ but breaks $\widetilde{\mathcal{C}}_{2y}$. As discussed in Sec.~\ref{subsec_hg_Dirac}, $\widetilde{\mathcal{C}}_{2y}$ causes the presence of the Dirac point at $Y$ instead of the Weyl loop, and enforces the double degeneracy on $\Gamma$-$Y$ and $Y$-$M$. Now, since the shear strain breaks $\widetilde{C}_{2y}$, the double degeneracy on $\Gamma$-$Y$ and $Y$-$M$ is expected to split into a hourglass dispersion, and the Dirac point at $Y$ should evolve into an hourglass Weyl loop. As a result, one expects two hourglass Weyl loops in shear strained monolayer GaTeI, one around $X$ and the other around $Y$. To confirm this, we apply shear strain to the GaTeI monolayer, by changing the angle $\gamma$ between $a$ and $b$ axis. As shown in Fig.~\ref{fig_shear}(d), one indeed observes a second hourglass Weyl loop around $Y$. This suggests that the shear strain can be employed as an effective way to control the number of hourglass Weyl loops in monolayer GaTeI.

\subsection{Spin-Orbit Weyl point}
\label{sec_hgWeyl_GaTeI}

\begin{figure}
	\includegraphics[width=0.45\textwidth]{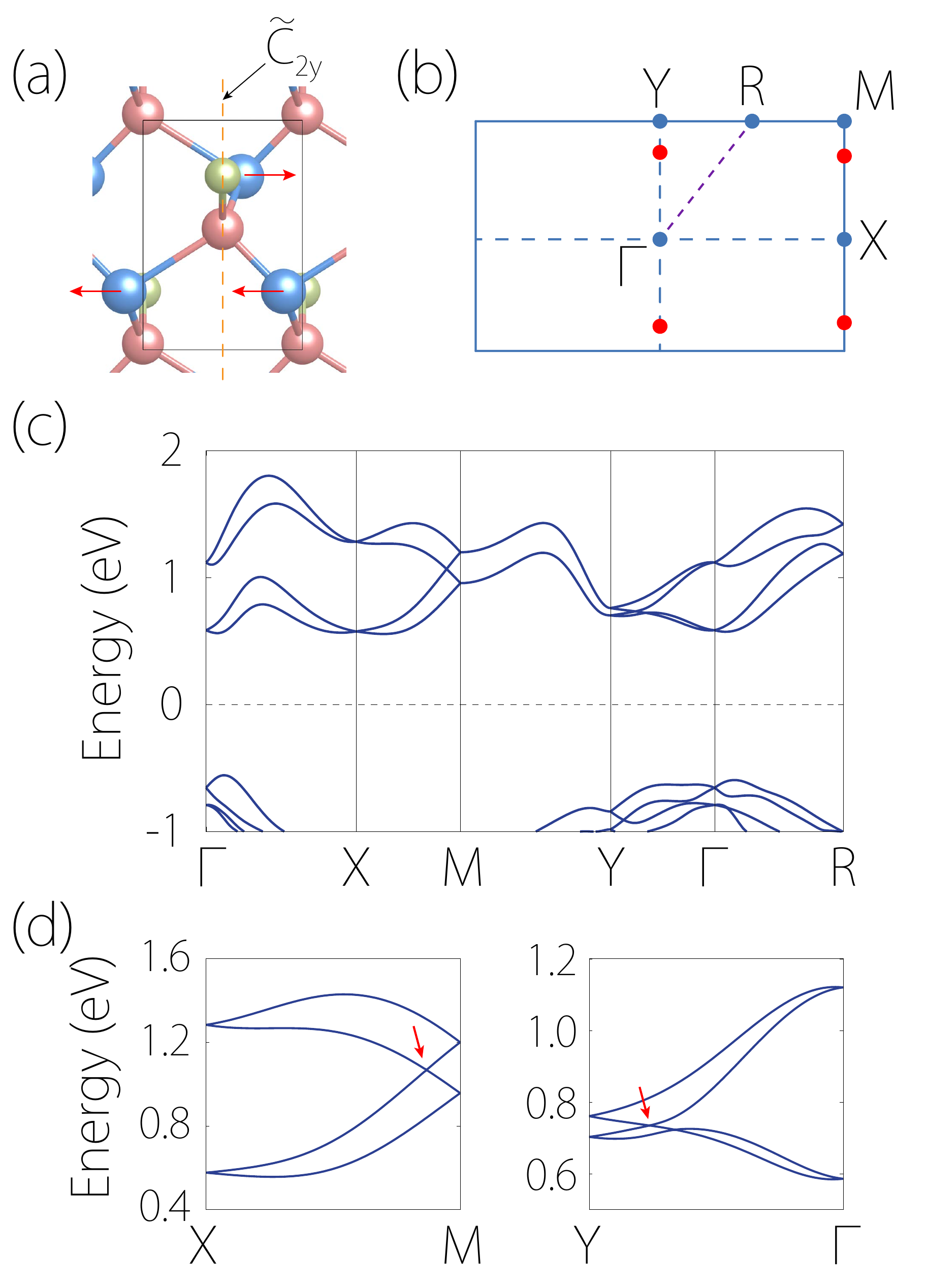}
	\caption{(a) Breaking the glide mirror symmetry by shifting one of the Te atoms in the unit cell along the $+a$ direction, while shifting the other Te atom along the $-a$ direction. The screw axis is preserved. (b) Schematic figure for the 2D Brillouin zone. The red dots indicate the emergent spin-orbit Weyl points on $\Gamma$-$Y$ and $X$-$M$. $R$ is the midpoint between $M$ and $Y$. (c) Calculated electronic band structure with SOC included. Panel (d) shows the zoom-in images for the bands along $\Gamma$-$Y$ and $X$-$M$, showing the hourglass-type dispersions. The red arrows indicate the (isolated) spin-orbit Weyl points.}
	\label{fig_hgWeyl}
\end{figure}

As discussed above, the hourglass Weyl loops in 2D requires the presence of the nonsymmorphic glide mirror symmetry $\widetilde{\mathcal{M}}_{z}$. If $\widetilde{\mathcal{M}}_{z}$ is violated, then the Weyl loop should be destroyed. For example, a vertical applied electric field can gap out the original Weyl loop.

Another interesting possibility is that the Weyl loop is \emph{partially} gapped with certain Weyl points on the loop preserved. We explore such possibility in the monolayer GaTeI system. We find that if $\widetilde{\mathcal{M}}_{z}$ is broken but $\widetilde{\mathcal{C}}_{2y}$ is preserved, it would result in two pairs of spin-orbit Weyl points on the two sides of $Y$ and $M$ along the $k_{x}=0$ and $k_{x}=\pi$ lines, respectively. These Weyl points are essential, dictated by $\mathcal{T}$ and $\widetilde{\mathcal{C}}_{2y}$ symmetries, as we discuss below.

For the twofold screw axis $\widetilde{\mathcal{C}}_{2y}$:~$(x,y,z)\rightarrow(-x+\frac{1}{2},y+\frac{1}{2},-z)$, in the presence of SOC, one finds that
\begin{equation}
  (\widetilde{\mathcal{C}}_{2y})^{2}=T_{01}\bar{E}=-e^{-ik_{y}}.
\end{equation}
The eigenvalues of $\widetilde{\mathcal{C}}_{2y}$ are therefore given by $s_{y}=\pm ie^{-ik_{y}/2}$. Consider the line $k_{x}=0$, on which the states can be chosen as the eigenstates of $\widetilde{\mathcal{C}}_{2y}$. One notes that at the TRIM point $\Gamma$, since $s_{y}=\pm i$, each Kramers pair $|u\rangle$ and $\mathcal{T}|u\rangle$ at $\Gamma$ must have opposite $s_{y}$ (one with $s_{y}=+i$ and the other with $s_{y}=-i$), while at the TRIM point $Y$ ($0,\pi$), $s_{y}=\pm 1$, such that each Kramers pair $|u\rangle$ and $\mathcal{T}|u\rangle$ at $Y$ should have the same $s_{y}$ ($+1$ or $-1$). Thus, there must be a partner switching when going from $\Gamma$ to $Y$, leading to the hourglass-type dispersion, which shares a similar pattern as in Fig.~\ref{fig_hourglass}(a). The doubly degenerate neck crossing point is protected since the two crossing bands have opposite $s_{y}$. This degeneracy is generally lifted away from this line due to the loss of symmetry protection, so that an isolated Weyl point is formed on $\Gamma$-$Y$. Due to the time reversal symmetry, we should have a pair of such Weyl points on the $k_{x}=0$ line. Similarly, one can show that another pair exists on the $k_{x}=\pi$ line.

To verify our above argument, we consider the following perturbations to the lattice structure of monolayer GaTeI. In order to break $\widetilde{\mathcal{M}}_{z}$ while preserving $\widetilde{\mathcal{C}}_{2y}$, we artificially shift one of the Te atoms in a unit cell along the $+a$ direction (by $0.5~\text{\AA}$), and shift the other Te atom along the $-a$ direction (also by $0.5~\text{\AA}$), as shown in Fig.~\ref{fig_hgWeyl}(a). As a result, we find that the original hourglass Weyl loop around $X$ is indeed destroyed, while the hourglass-type dispersion along $X$-$M$ ($k_{x}=\pi$ line) is preserved. Meanwhile, the double degeneracy on $\Gamma$-$Y$ is removed, with four bands entangled to form another hourglass, as shown in Fig.~\ref{fig_hgWeyl}(d). We also check the band structure along a generic path $\Gamma$-$R$, on which no crossing points is observed. This verifies that the original Weyl loop gives way to isolated Weyl points under the symmetry breaking. The distribution of the emerging Weyl points in the BZ is schematically illustrated in Fig.~\ref{fig_hgWeyl}(b).

We stress that the spin-orbit Weyl point here is distinct from other nodal points (in 2D materials) discussed before. It is distinct from the spin-orbit Dirac point (as in monolayer HfGeTe~\cite{Guan2017Two-PRM}) in terms of the degeneracy. The spin-orbit Weyl point is doubly degenerate, whereas the spin-orbit Dirac point is fourfold degenerate. It is also distinct from the spin-orbit-free Weyl point (as in graphene)~\footnote{Due to historical reasons, spin-orbit-free Weyl points in 2D are also widely
referred to as Dirac points. See the discussion in Ref.~\cite{Yang2016Dirac-spin}.}, because it is stable under SOC. Thus, the spin-orbit Weyl point represents a new type of 2D nodal structure, which should be further explored in future works.

\section{DISCUSSION AND CONCLUSION}
\label{discussion}
In this work, we have presented symmetry conditions for realizing hourglass Weyl loops in 2D systems, and we have found the first realistic material platform. The existence of such Weyl loops is solely dictated by the nonsymmorphic space group symmetry, so that the features discussed for monolayer GaTeI are also shared by other members of the family (see Appendix~\ref{asec_materials} for the band structure results). Generally, the Weyl-loop features are better resolved for members with heavier elements (which have a stronger SOC). In addition, the analysis here can be directly applied for systems with similar symmetries, especially for those 2D materials with the space group No.~31.

However, it should be noted that symmetry cannot constrain the energy of the band crossings. The material examples presented here are still not ideal in the sense that the Weyl loops/points here are not very close to the Fermi level. Nevertheless, they serve the purpose to demonstrate that these novel band crossings can indeed appear in realistic 2D systems and to study their interesting transformations.  The results presented here pave the way to search for more candidate materials with such novel emergent fermions in future studies.

In addition, the interesting band features may be moved closer to Fermi level by doping, strain, or pressure engineering. For monolayer GaTeI, the conduction band states can be access by electron doping. For 2D materials, the electron doping techniques are under rapid development, and efficient carrier doping has been demonstrated by ion liquid gating~\cite{Mak2013Tightly-Nm,Zhang2014Electrically-S}. The band crossing features can be imaged by using the angle-resolved photoemission spectroscopy (ARPES). For unoccupied states, this can be achieved by using a pump-probe setup with the time-resolved ARPES technique~\cite{Schmitt2008Transient-S,Sobota2012Ultrafast-Prl}, i.e., after being pumped by the first laser, the excited electrons are probed by ARPES with a second laser to map out their energy and momentum distribution. Besides, the conduction band states may also be probed by the scanning tunneling spectroscopy on the quasiparticle interference pattern~\cite{Zheng2016Atomic-An,Zhu2018Quasiparticle-Nc}.

In conclusion, we have theoretically investigated the hourglass Weyl loops in 2D systems. We present symmetry conditions for realizing the hourglass Weyl loops in 2D, which involve the presence of nonsymmorphic lattice symmetry. We find that these loops can exhibit different patterns and topologies in the BZ. We identify the monolayer GaTeI family materials as realistic examples to host hourglass Weyl loops.
We show that there is one loop around $X$ point, while the other loop around $Y$ shrinks to a spin-orbit Dirac point due to the presence of an additional screw axis. Interestingly, if the glide mirror is violated while the screw axis is preserved, the hourglass Weyl loop and the spin-orbit Dirac point would transform into spin-orbit Weyl points. The spin-orbit Weyl loops and the spin-orbit Weyl points are new band crossing features distinct from those found in 2D materials before.
Our findings offer useful guidance for the material search and identify a concrete material platform to explore the intriguing physics of these topological band structures.

\begin{acknowledgments}
The authors thank B. Tai, J.-M. Ma and D. L. Deng for helpful discussions. This work is supported by Singapore Ministry of Education AcRF Tier 2 (Grant No. MOE2017-T2-2-108). We
acknowledge computational support from the Texas Advanced Computing Center and the National Supercomputing Centre Singapore.
\end{acknowledgments}

\begin{appendix}
	
\renewcommand{\theequation}{A\arabic{equation}}
\setcounter{equation}{0}
\renewcommand{\thefigure}{A\arabic{figure}}
\setcounter{figure}{0}
\renewcommand{\thetable}{A\arabic{table}}
\setcounter{table}{0}

\section{Details of first-principles calculations}\label{asec_DFT}
Our first-principle calculations are based on the density functional theory (DFT), as implemented in the Vienna \emph{ab initio} simulation package~\cite{Kresse1993Ab-PRB,Kresse1996Efficient-PRB}. The projector augmented wave method was adopted~\cite{Bloechl1994Projector-PRB}. The generalized gradient approximation (GGA) with the Perdew-Burke-Ernzerhof (PBE)~\cite{Perdew1996Generalized-PRL} realization was adopted for the exchange-correlation potential. Our main results are also verified by the hybrid functional approach (HSE06)~\cite{Heyd2003Hybrid-TJocp,Krukau2006Influence-TJocp}. The structures are fully optimized with the energy and force convergence criteria of $10^{-6}$~eV and $10^{-2}$~eV/\text{\AA}, respectively. The plane-wave energy cutoff is set to be 350 eV, and the BZ was sampled with $\Gamma$-centered $k$ mesh of size $13 \times 9 \times 4$ for the 3D bulk and $12 \times 9 \times 1$ for monolayer. The optimized van der Waals (vdW) correlation functional optB86b-vdW has been taken in to account in the exfoliation energy calculation and the bulk calculation~\cite{Klimevs2011Van-PRB}. A vacuum layer of $15~\text{\AA}$ thickness is added to avoid artificial interactions between periodic images for monolayer calculations. The phonon spectrum is obtained by using the PHONOPY code~\cite{Togo2015First-SM}, based on the force constants calculated by the VASP-DFPT method.


\section{Band structure without SOC}
\label{asec_materials}

\begin{figure}
	\centering
	\includegraphics[width=0.4\textwidth]{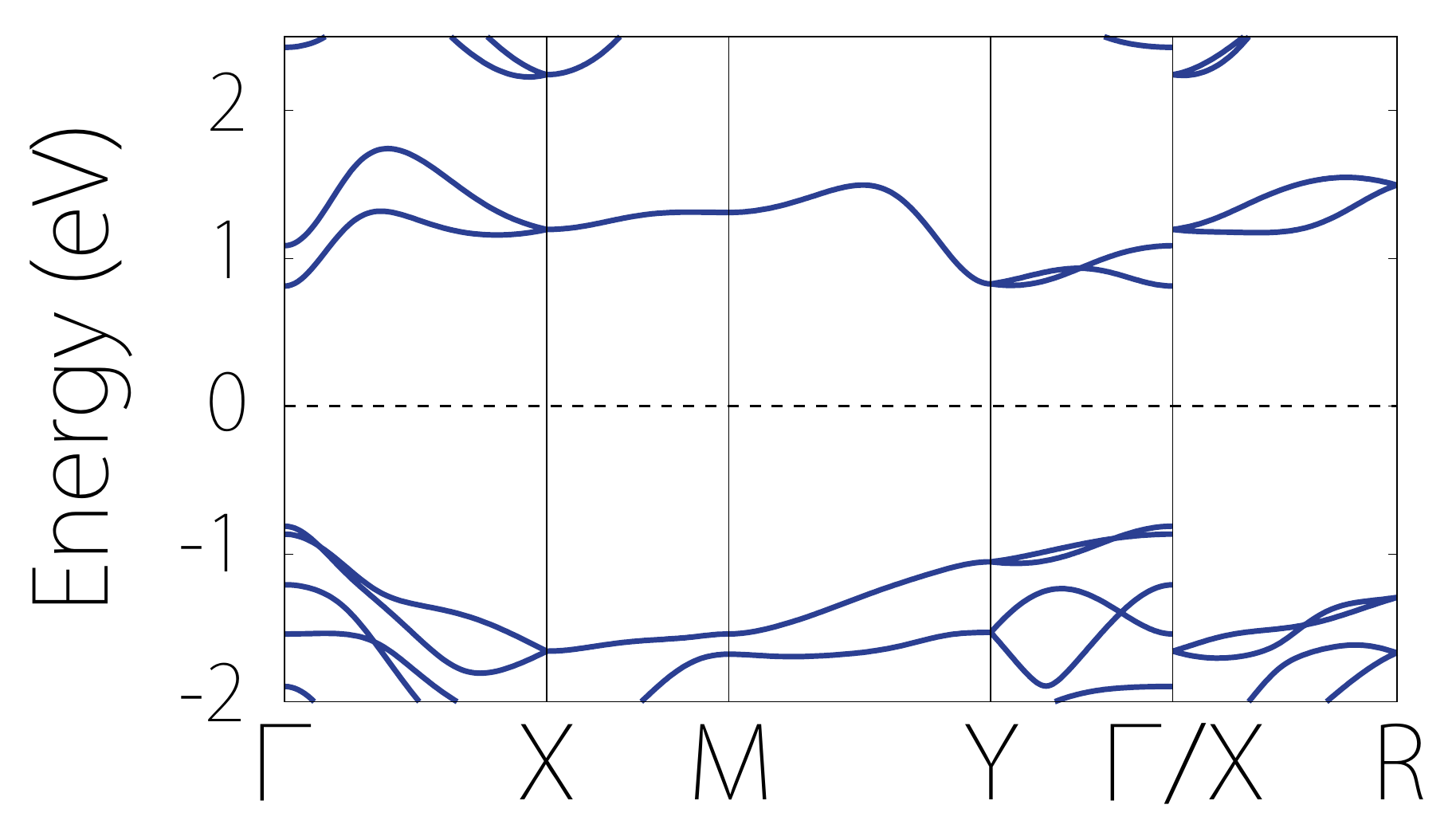}
	\caption{Calculated band structure for monolayer GaTeI in the absence of SOC}
	\label{fig_bands_noSOC}
\end{figure}
In the main text, we have discussed the electronic band structure of monolayer GaTeI with SOC included. Here, we consider its band structure in the absence of SOC.  The result is shown in Fig.~\ref{fig_bands_noSOC}. One observes that the GaTeI monolayer is also a semiconductor but has a direct band gap of 1.62~eV. For the bands near CBM, it shows a twofold degeneracy on the BZ boundary, such as the $X$-$M$-$Y$ path. The double degeneracy along $M$-$Y$ is a result of the Kramers-like degeneracy due to the antiunitary $\mathcal{T}\widetilde{\mathcal{C}}_{2y}$ symmetry, while the double degeneracy on $X$-$M$ results from the relation $\widetilde{\mathcal{M}}_{z}\widetilde{\mathcal{C}}_{2y} = -\widetilde{\mathcal{C}}_{2y}\widetilde{\mathcal{M}}_{z}$ on this path. The symmetry analysis is similar to those in the presence of SOC (see Appendices~\ref{asec_WeylGY} and~\ref{asec_WeylYM}).

\section{Band structure from hybrid functional method}\label{asec_hse}
\begin{figure}
	\centering
	\includegraphics[width=0.4\textwidth]{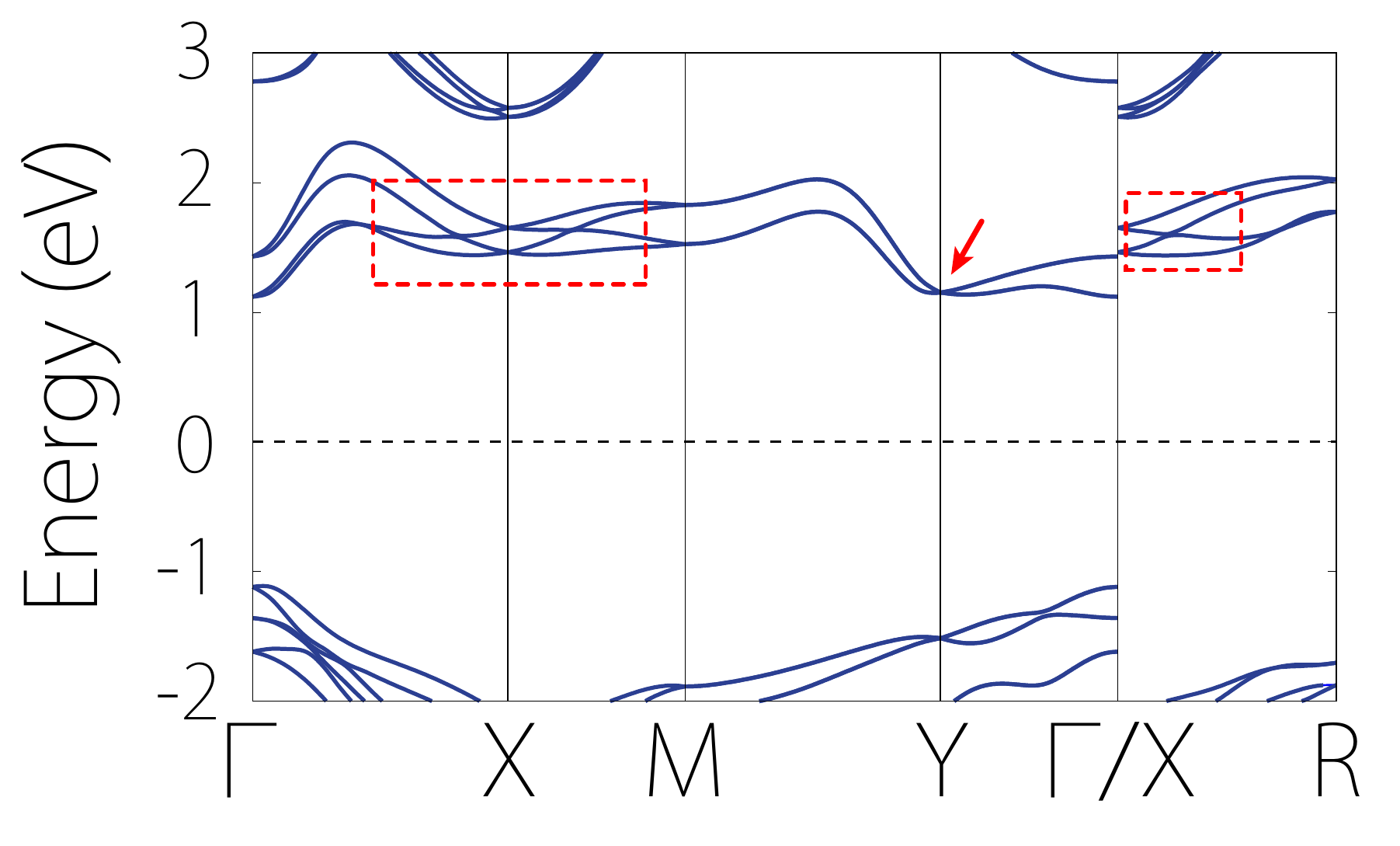}
	\caption{{Electronic band structure of monolayer GaTeI calculated by hybrid functional (HSE06) method. The hourglass Weyl loop around $X$ is indicated by a red dashed box, while the Dirac point at $Y$ is indicated by a red arrow.}}
	\label{fig_bands_hse}
\end{figure}

Figure~\ref{fig_bands_hse} shows the band structure of monolayer GaTeI obtained by the hybrid functional (HSE06) method. Compared with the GGA result, one can observe that all the topological band features are maintained. The main difference is that the gap size is increased to about 2.2~eV.

\section{Double degeneracy on $\Gamma$-$Y$ path}\label{asec_WeylGY}
Here, we demonstrate the double degeneracy on the $\Gamma$-$Y$ path in the presence of SOC. The $\Gamma$-$Y$ path is an invariant subspace of both $\widetilde{\mathcal{C}}_{2y}$ and $\widetilde{\mathcal{M}}_{z}$. The two operations satisfy the following algebra
\begin{equation}
\widetilde{\mathcal{M}}_{z}\widetilde{\mathcal{C}}_{2y}=-T_{10}\widetilde{\mathcal{C}}_{2y}\widetilde{\mathcal{M}}_{z}=-e^{-ik_{x}}
\widetilde{\mathcal{C}}_{2y}\widetilde{\mathcal{M}}_{z},
\end{equation}
where the minus sign is due to the anticommutativity between two spin rotations, i.e., $\{\sigma_{z},\sigma_{y}\}=0$, so that $\{\widetilde{\mathcal{M}}_{z},\widetilde{\mathcal{C}}_{2y}\}=0$ on $\Gamma$-$Y$. Consequently, for an eigenstate $|g_{z}\rangle$ of $\widetilde{\mathcal{M}}_{z}$ with eigenvalue $g_{z}$, the following relation holds:
\begin{equation}
\widetilde{\mathcal{M}}_{z}(\widetilde{\mathcal{C}}_{2y}|g_{z}\rangle)=-g_{z}(\widetilde{\mathcal{C}}_{2y}|g_{z}\rangle),
\end{equation}
showing that $|g_{z}\rangle$ and $\widetilde{\mathcal{C}}_{2y}|g_{z}\rangle$ are distinct states degenerate at the same energy. This ensures the double degeneracy along $\Gamma$-$Y$.

\section{Double degeneracy on $Y$-$M$ path}\label{asec_WeylYM}
Here, we present the detailed analysis of the double degeneracy on $Y$-$M$ in the presence of SOC. In the presence of SOC, we have $\mathcal{T}^{2}=-1$ and $(\widetilde{\mathcal{C}}_{2y})^{2}=T_{01}\bar{E}=-e^{-ik_{y}}$. Since $[\mathcal{T},\widetilde{\mathcal{C}}_{2y}]=0$, the combined operation $\mathcal{T}\widetilde{\mathcal{C}}_{2y}$ satisfies
\begin{equation}
(\mathcal{T}\widetilde{\mathcal{C}}_{2y})^{2}=e^{-ik_{y}}.
\end{equation}
One notes that any point on the $Y$-$M$ path is invariant under $\mathcal{T}\widetilde{\mathcal{C}}_{2y}$. From the above equation, the antiunitary symmetry satisfies
\begin{equation}
(\mathcal{T}\widetilde{\mathcal{C}}_{2y})^{2}=-1
\end{equation}
on $Y$-$M$. Thus, the Kramers double degeneracy arises at each point on this path. The above argument is similar to that for the Class-II nodal surface in 3D systems discussed in the previous work~\cite{Wu2018Nodal-PRB}.

\section{Results for Other Members of the Family}

\begin{figure*}
	\centering
	\includegraphics[width=0.8\textwidth]{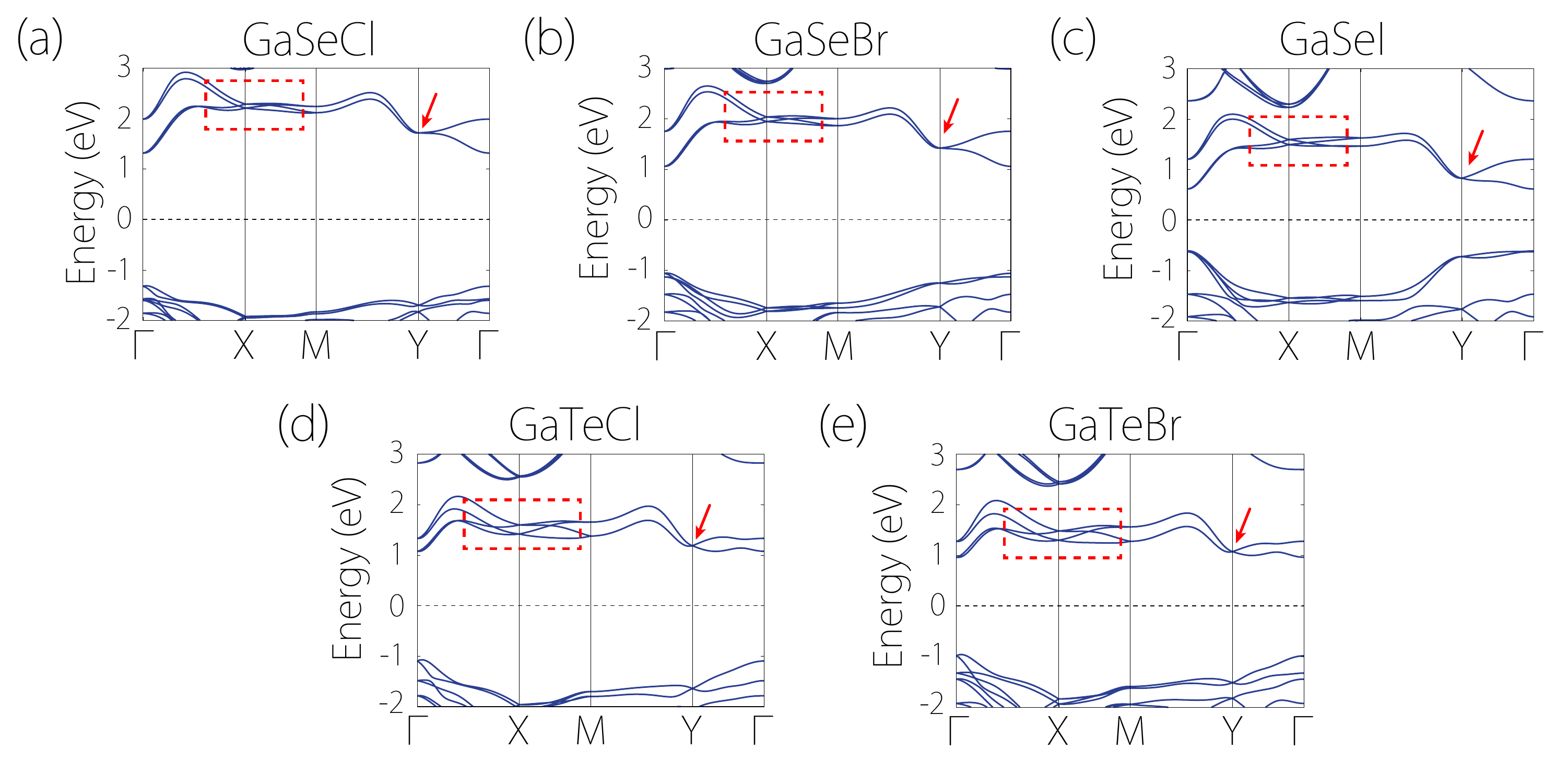}
	\caption{Band structures of other monolayer materials in GaXY (X = Se, Te; Y = Cl, Br, I) compounds with SOC: (a) GaSeCl, (b) GaSeBr, (c) GaSeI, (d) GaTeCl, and (e) GaTeBr. The hourglass Weyl loop around $X$ is indicated by a red dashed box, and the Dirac point at $Y$ is indicated by a red arrow.}
	\label{fig_bands_other}
\end{figure*}

Figure~\ref{fig_bands_other} shows the calculated band structures of some other members of the monolayer GaTeI material family. One can see that they also possess the hourglass Weyl loop around $X$ as well as the Dirac point at $Y$, as discussed in the main text.

\section{Lattice model for hourglass Weyl loop}\label{asec_tb}
\begin{figure}
	\centering
	\includegraphics[width=0.5\textwidth]{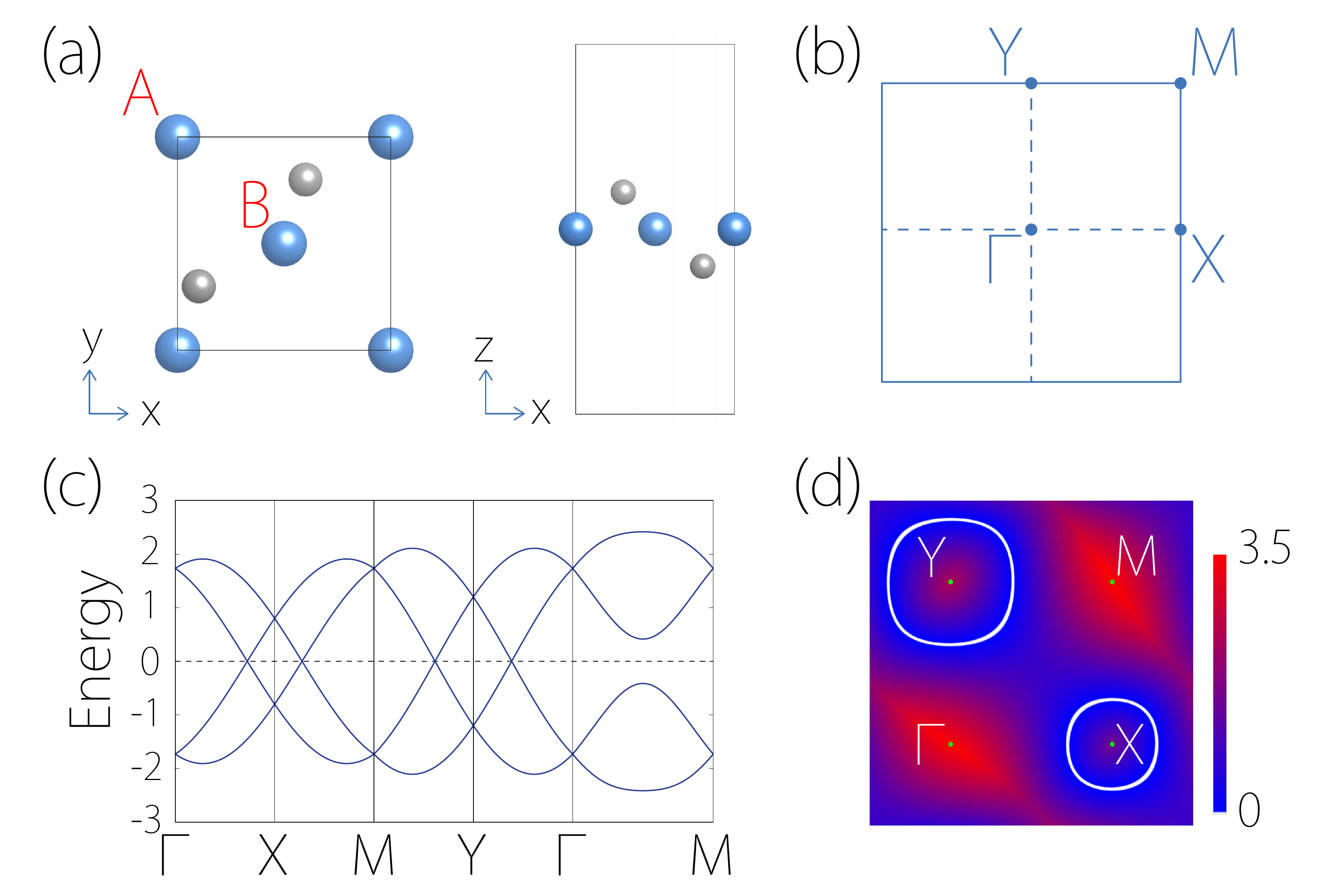}
	\caption{(a) Top and side views of the square lattice for our lattice model. It possesses only the symmetries $\widetilde{M}_z=\{M_{z}|\frac{1}{2}\frac{1}{2}\}$ and $\mathcal{T}$. The unit cell contains two active sites labeled as $A$ and $B$ (denoted by blue balls). The gray-colored balls denote inactive sites that do not directly enter the model, but enforce the symmetry condition by affecting the hopping amplitudes between the active sites. (b) The corresponding Brillouin zone. (c) A typical band structure for the model. The parameters are taken as $t_{2}=-1.0, t_{1}^\text{SO} = 1.0, t_{2}^\text{SO} = -0.2, t_{3}^\text{SO} = -1.0$, and $t_{6}^\text{SO} = -1.0$, while the remaining ones are set to $0$. (d) Shape of the two hourglass Weyl loops.}
	\label{fig_tb}
\end{figure}
Here, we construct a minimal lattice model for the hourglass Weyl loop. To be specific, we consider Case-I discussed in Sec.~\ref{sec_symm_hg_loop}. The relevant symmetries are the glide mirror $\widetilde{\mathcal{M}}_z=\{M_{z}|\frac{1}{2}\frac{1}{2}\}$ and the $\mathcal{T}$, which corresponds to space group No.~7. The simplest setup is a 2D square lattice, with each unit cell containing two active sites ($A$ and $B$) as illustrated in Fig.~\ref{fig_tb}(a). Each active (blue-colored) site has an $s$-like orbital with two spin states, such that we totally have four bands. In the basis $\{|A, \uparrow\rangle, |A, \downarrow\rangle, |B, \uparrow\rangle, |B, \downarrow\rangle \}$, the symmetry operations can be represented as
\begin{equation}
\begin{aligned}
	\widetilde{\mathcal{M}}_{z} = -i\tau_{x}\sigma_{z},\qquad
	\mathcal{T} = -i\sigma_{y}\mathcal{K},
\end{aligned}
\end{equation}
where the Pauli matrices $\tau_{i}$ and $\sigma_{i}$ act on the sublattice and the spin spaces, respectively, and $\mathcal{K}$ is the complex conjugation operation. Then the symmetry allowed lattice Hamiltonian up to the first-neighbor hopping can be obtained as
\begin{widetext}
\begin{equation}
\begin{aligned}
    \mathcal{H} =&~\varepsilon_{0} + (t_{1} \cos{\frac{k_{x}+k_{y}}{2}}+t_{2}\cos{\frac{k_{x}-k_{y}}{2}})\tau_{x}+ (t_{1}^\text{SO} \sin{\frac{k_{x}+k_{y}}{2}}+t_{2}^\text{SO}\sin{\frac{k_{x}-k_{y}}{2}})\tau_{x}\sigma_{z} \\
	&+ (t_{3}^\text{SO} \cos{\frac{k_{x}+k_{y}}{2}}+t_{4}^\text{SO}\cos{\frac{k_{x}-k_{y}}{2}})\tau_{y}\sigma_{x} + (t_{5}^\text{SO} \cos{\frac{k_{x}+k_{y}}{2}}+t_{6}^\text{SO}\cos{\frac{k_{x}-k_{y}}{2}})\tau_{y}\sigma_{y},
\end{aligned}
\end{equation}
\end{widetext}
where the coefficients $t_{i}$ and $t_{i}^\text{SO}$ are real valued model parameters, $\sigma$'s and $\tau$'s are the Pauli matrices. Figure~\ref{fig_tb}(c) shows a typical band structure obtained for this model, which indeed exhibits a pair of hourglass Weyl loops circling around $X$ and $Y$ [Fig.~\ref{fig_tb}(d)], consistent with the pattern in Fig.~\ref{fig_hourglass}(b).

\end{appendix}


%

\end{document}